\documentclass[twocolumn,aps,showpacs,superscriptaddress]{revtex4-1}
\usepackage{graphicx}
\usepackage{dcolumn}
\usepackage{bm}
\usepackage[pagewise,mathlines]{lineno}
\usepackage{color}
\usepackage{ulem}
\usepackage[none]{hyphenat}

\newcommand{ \be }{\begin{equation}}
\newcommand{ \ee }{\end{equation}}
\newcommand{ \bea }{\begin{eqnarray}}
\newcommand{ \eea }{\end{eqnarray}}
\newcommand{ \bff }{\begin{figure}[htpb]}
\newcommand{ \ef }{\end{figure}}
\newcommand{ \bmn }{\begin{minipage}}
\newcommand{ \emn }{\end{minipage}}
\newcommand{ \bt }{\begin{table}[htpb]}
\newcommand{ \et }{\end{table}}

\newcommand{ \pt }{$p_{T}$}

\newcommand{\epem}{$e^{+}e^{-}$}

\usepackage{amssymb}

\begin{document}
\def\Journal#1#2#3#4{{#1} {\bf #2}, #3 (#4)}

\def\NCA{Nuovo Cimento}
\def\NIM{Nucl. Instr. Meth.}
\def\NIMA{{Nucl. Instr. Meth.} A}
\def\NPB{{Nucl. Phys.} B}
\def\NPA{{Nucl. Phys.} A}
\def\PLB{{Phys. Lett.}  B}
\def\PRL{Phys. Rev. Lett.}
\def\PRC{{Phys. Rev.} C}
\def\PRD{{Phys. Rev.} D}
\def\ZPC{{Z. Phys.} C}
\def\JPG{{J. Phys.} G}
\def\CPC{Comput. Phys. Commun.}
\def\EPJ{{Eur. Phys. J.} C}
\def\PR{Phys. Rept.}
\def\JHEP{JHEP}



\preprint{}
\title{Direct virtual photon production in Au+Au collisions at $\sqrt{s_{NN}}$ = 200 GeV}


\affiliation{AGH University of Science and Technology, FPACS, Cracow 30-059, Poland}
\affiliation{Argonne National Laboratory, Argonne, Illinois 60439}
\affiliation{Brookhaven National Laboratory, Upton, New York 11973}
\affiliation{University of California, Berkeley, California 94720}
\affiliation{University of California, Davis, California 95616}
\affiliation{University of California, Los Angeles, California 90095}
\affiliation{Central China Normal University, Wuhan, Hubei 430079}
\affiliation{University of Illinois at Chicago, Chicago, Illinois 60607}
\affiliation{Creighton University, Omaha, Nebraska 68178}
\affiliation{Czech Technical University in Prague, FNSPE, Prague, 115 19, Czech Republic}
\affiliation{Nuclear Physics Institute AS CR, 250 68 Prague, Czech Republic}
\affiliation{Frankfurt Institute for Advanced Studies FIAS, Frankfurt 60438, Germany}
\affiliation{Institute of Physics, Bhubaneswar 751005, India}
\affiliation{Indiana University, Bloomington, Indiana 47408}
\affiliation{Alikhanov Institute for Theoretical and Experimental Physics, Moscow 117218, Russia}
\affiliation{University of Jammu, Jammu 180001, India}
\affiliation{Joint Institute for Nuclear Research, Dubna, 141 980, Russia}
\affiliation{Kent State University, Kent, Ohio 44242}
\affiliation{University of Kentucky, Lexington, Kentucky, 40506-0055}
\affiliation{Lamar University, Physics Department, Beaumont, Texas 77710}
\affiliation{Institute of Modern Physics, Chinese Academy of Sciences, Lanzhou, Gansu 730000}
\affiliation{Lawrence Berkeley National Laboratory, Berkeley, California 94720}
\affiliation{Lehigh University, Bethlehem, PA, 18015}
\affiliation{Max-Planck-Institut fur Physik, Munich 80805, Germany}
\affiliation{Michigan State University, East Lansing, Michigan 48824}
\affiliation{National Research Nuclear University MEPhI, Moscow 115409, Russia}
\affiliation{National Institute of Science Education and Research, Bhubaneswar 751005, India}
\affiliation{National Cheng Kung University, Tainan 70101 }
\affiliation{Ohio State University, Columbus, Ohio 43210}
\affiliation{Institute of Nuclear Physics PAN, Cracow 31-342, Poland}
\affiliation{Panjab University, Chandigarh 160014, India}
\affiliation{Pennsylvania State University, University Park, Pennsylvania 16802}
\affiliation{Institute of High Energy Physics, Protvino 142281, Russia}
\affiliation{Purdue University, West Lafayette, Indiana 47907}
\affiliation{Pusan National University, Pusan 46241, Korea}
\affiliation{Rice University, Houston, Texas 77251}
\affiliation{University of Science and Technology of China, Hefei, Anhui 230026}
\affiliation{Shandong University, Jinan, Shandong 250100}
\affiliation{Shanghai Institute of Applied Physics, Chinese Academy of Sciences, Shanghai 201800}
\affiliation{State University Of New York, Stony Brook, NY 11794}
\affiliation{Temple University, Philadelphia, Pennsylvania 19122}
\affiliation{Texas A\&M University, College Station, Texas 77843}
\affiliation{University of Texas, Austin, Texas 78712}
\affiliation{University of Houston, Houston, Texas 77204}
\affiliation{Tsinghua University, Beijing 100084}
\affiliation{University of Tsukuba, Tsukuba, Ibaraki, Japan,}
\affiliation{Southern Connecticut State University, New Haven, CT, 06515}
\affiliation{United States Naval Academy, Annapolis, Maryland, 21402}
\affiliation{Valparaiso University, Valparaiso, Indiana 46383}
\affiliation{Variable Energy Cyclotron Centre, Kolkata 700064, India}
\affiliation{Warsaw University of Technology, Warsaw 00-661, Poland}
\affiliation{Wayne State University, Detroit, Michigan 48201}
\affiliation{World Laboratory for Cosmology and Particle Physics (WLCAPP), Cairo 11571, Egypt}
\affiliation{Yale University, New Haven, Connecticut 06520}

\author{L.~Adamczyk}\affiliation{AGH University of Science and Technology, FPACS, Cracow 30-059, Poland}
\author{J.~K.~Adkins}\affiliation{University of Kentucky, Lexington, Kentucky, 40506-0055}
\author{G.~Agakishiev}\affiliation{Joint Institute for Nuclear Research, Dubna, 141 980, Russia}
\author{M.~M.~Aggarwal}\affiliation{Panjab University, Chandigarh 160014, India}
\author{Z.~Ahammed}\affiliation{Variable Energy Cyclotron Centre, Kolkata 700064, India}
\author{N.~N.~Ajitanand}\affiliation{State University Of New York, Stony Brook, NY 11794}
\author{I.~Alekseev}\affiliation{Alikhanov Institute for Theoretical and Experimental Physics, Moscow 117218, Russia}\affiliation{National Research Nuclear University MEPhI, Moscow 115409, Russia}
\author{D.~M.~Anderson}\affiliation{Texas A\&M University, College Station, Texas 77843}
\author{R.~Aoyama}\affiliation{University of Tsukuba, Tsukuba, Ibaraki, Japan,}
\author{A.~Aparin}\affiliation{Joint Institute for Nuclear Research, Dubna, 141 980, Russia}
% ...

\collaboration{STAR Collaboration}\noaffiliation

\date{\today}

\begin{abstract}
We report the direct virtual photon invariant yields in the transverse momentum ranges $1\!<\!p_{T}\!<\!3$ GeV/$c$ and $5\!<\!p_T\!<\!10$ GeV/$c$ at mid-rapidity derived from the dielectron invariant mass continuum region $0.10<M_{ee}<0.28$ GeV/$c^{2}$ for 0-80\% minimum-bias Au+Au collisions at $\sqrt{s_{NN}}=200$ GeV. A clear excess in the invariant yield compared to the nuclear overlap function $T_{AA}$ 
scaled $p+p$ reference is observed in the \pt~range $1\!<\!p_{T}\!<\!3$ GeV/$c$. For $p_T\!>6$ GeV/$c$ the production follows $T_{AA}$ scaling. Model calculations with contributions from thermal radiation and initial hard parton scattering are consistent within uncertainties with the direct virtual photon invariant yield.
\end{abstract}



\maketitle

\section{Introduction}
\label{intro}
Photon production provides a unique observable to study the fundamental properties of the hot and dense medium created in ultra-relativistic heavy-ion collisions. They are produced during all stages of the collisions and from all forms of the created matter. Due to minimal interactions with this matter, photons can convey information about the dynamics of the entire time evolution of the medium~\cite{dileptonII}. Direct photons are defined to be all produced photons except those from hadron decays in the last stage of the collision. They include photons produced in the initial stage through hard scattering, those from thermal radiation, which are photons radiated from the thermally equilibrated partons and hadrons, fragmentation photons, and those from jet-plasma interactions. Measurements at RHIC~\cite{phenixgamma} and the LHC~\cite{DVP_CMS,ALICEphoton} have shown that the production of high $p_T$ direct photons in heavy-ion collisions is consistent with the $p+p$ result scaled by the nuclear overlap function $T_{AA}$ for $p_T\!>\!$ 5 GeV/$c$. These results indicate that high $p_T$ production is dominated by hard processes.

At $1\!<p_{T}\!<3$ GeV/$c$ thermal contributions from the hadronic medium and Quark-Gluon Plasma (QGP) play a major role~\cite{Gale:08}. At $3\!<\!p_{T}\!<\!5$ GeV/$c$ the interaction of high energy partons with the QGP (e.g. $q + g \rightarrow \gamma + q $) has been predicted to contribute a major part of the direct photon production~\cite{Gale:08}. An excess of direct photon yields compared to the $T_{AA}$ scaled $p+p$ production was found in central Au+Au at $\sqrt{s_{_{NN}}} = 200$ GeV in the $p_T$ range $0.4\!<\!p_T\!<4.0$ GeV/$c$~\cite{thermalphoton,Mizuno:14}  and in central Pb+Pb at $\sqrt{s_{_{NN}}} = 2.76$ TeV for $0.9\!<\!p_T\!<2.1$ GeV/$c$~\cite{ALICEphoton}. The excess increases exponentially as $p_T$ decreases. Moreover, the azimuthal anisotropy ($v_2$) of direct photons has been found to be substantial in the range $1\!<\!p_{T}\!<\!4$ GeV/$c$ in 0-20\% central Au+Au collisions at $\sqrt{s_{_{NN}}} = 200$ GeV~\cite{photonv2}.

Model calculations~\cite{rapp:11,shen:14} including QGP and hadronic medium thermal photons describe the excess yields in Pb+Pb collisions at $\sqrt{s_{_{NN}}} = 2.76$ TeV reasonably well, but fail to simultaneously describe the excess yields and large $v_2$ observed in Au+Au at $\sqrt{s_{_{NN}}} = 200$ GeV. This calls for new ingredients in the theoretical model calculations and new measurements from various experiments which will provide different systematics and may shed light on the origin of direct photons in this kinematic region.

There are two methods for measuring direct photons. One is the real photon method in which one measures all inclusive photons and then subtracts the photons from hadron decays. The other one, used in this article, is the virtual photon method in which one measures virtual photons via their associated dielectron pairs ($\gamma^{*}~\to~e^{+}e^{-}$) and then deduces the direct photon from the relationship between virtual photon and direct photon yields~\cite{thermalphoton}. In the STAR experiment it is very challenging to measure direct photons for $1\!<\!p_{T}\!<\!3$ GeV/$c$ using the electromagnetic calorimeter due to limited detector granularity, large occupancy, and insufficient energy resolution. However, the STAR detector has excellent capabilities for measuring dielectrons both in $p+p$ and Au+Au collisions~\cite{starppdilepton,stardielectronauau,stardielectronauauPRC,stardielectronv2}. The STAR~\cite{star} Time-Of-Flight detector (TOF) with full azimuthal coverage~\cite{startof} along with a high rate data acquisition system allows direct virtual photon measurements down to $p_T$ of 1 GeV/$c$ in Au+Au collisions at $\sqrt{s_{_{NN}}} = 200$ GeV. These measurements will provide a direct comparison to the previous measurements in the same kinematic region, in order to address the above results for direct photons at $\sqrt{s_{_{NN}}}$ = 200 GeV. In this article, we report measurements of the dielectron continuum and derive the direct virtual photon invariant yields for $1\!<\!p_{T}\!<\!3$ GeV/$c$ and $5\!<\!p_T\!<\!10$ GeV/$c$. Comparisons to model calculations with thermal contributions from the hadronic medium and QGP are discussed.

\section{Experiment and Data Analysis}
\label{analysis}
The data used in this analysis are from Au+Au collisions at $\sqrt{s_{NN}}=200$ GeV collected by the STAR detector in year 2010 (run 10) and 2011 (run 11). There are 258 million and 488 million minimum-bias (0-80\%) events from run 10 and run 11, respectively, passing data quality assurance and vertex selection. The collision vertex is required to be within 30 cm of the mean of the vertex distribution along the beam line, nominally at the center of the Time Projection Chamber (TPC)~\cite{startpc}. In the plane perpendicular to the beam line, the collision vertex is selected within 2 cm of the beam line. To improve the measurement at high \pt, we also use 39 million events from run 11, triggered by the Barrel ElectroMagnetic Calorimeter (BEMC)~\cite{BEMC}, in which the transverse energy deposited
in a single tower, with a size of $\Delta \eta \times \Delta \phi = 0.05 \times 0.05$, is required to be larger than 4.3 GeV. These BEMC triggered
events correspond to 6.5 billion minimum-bias triggered events for the dielectron analysis at high $p_T$. The BEMC trigger significantly enhances the capability of STAR for high \pt~dielectron measurement.

The main subsystems used for electron identification are the TPC and the TOF for the minimum-bias and central triggered events. With the selection requirements on the particle energy loss ($dE/dx$) measured by the TPC~\cite{bichsel,pidpp08} and particle velocity ($\beta$) measured by the TOF~\cite{tofPID}, high purity electron samples were obtained~\cite{pidNIMA}. The electron purity (fraction of true electrons in the identified electron sample) is about 95\% in Au+Au minimum-bias collisions on average and is $p_{T}$ dependent from 0.2 to 2.0 GeV/$c$~\cite{stardielectronauauPRC,stardielectronv2}. The detailed cuts for electron identification are listed in Ref.~\cite{stardielectronauauPRC}. For BEMC-triggered events, the electron (positron) identification for $p_T^{e}\!>\!4.5$ GeV/$c$ uses a combination of TPC and BEMC information~\cite{NPE:11} where additional requirements on the ratio of momentum measured by the TPC to the energy deposited in the BEMC are utilized and required to be within 0.3 to 1.5. The electron identification for  $0.2\!<\!p_T^{e}\!<\!2.0$ GeV/$c$ utilizes the information from the TPC and TOF, in the same way as was done for the minimum-bias events. For $p_T^{e}\!>\!4.5$ GeV/$c$, a multiple-Gaussian function is used to fit the normalized $dE/dx$ distribution, with each Gaussian component representing a contribution from each particle species. The electron purity, obtained as in Ref.~\cite{stardielectronauauPRC}, is 78\% at $p_{T}^{e}=4.5$ GeV/$c$, decreases as $p_{T}^{e}$ increases, and reaches a value of 30\% at $p_{T}^{e}=10$ GeV/$c$. The electron purity as a function of $p_{T}^{e}$ for $4.5\!<\!p_T^{e}\!<\!10$ GeV/$c$ can be described by a fourth-order polynomial function $-6.57+4.69x-1.06x^2+0.10x^3-0.0034x^4$, in which $x=p^{e}_T$/(GeV/$c$).

\begin{figure}
\begin{center}
\includegraphics*[width=0.50\textwidth]{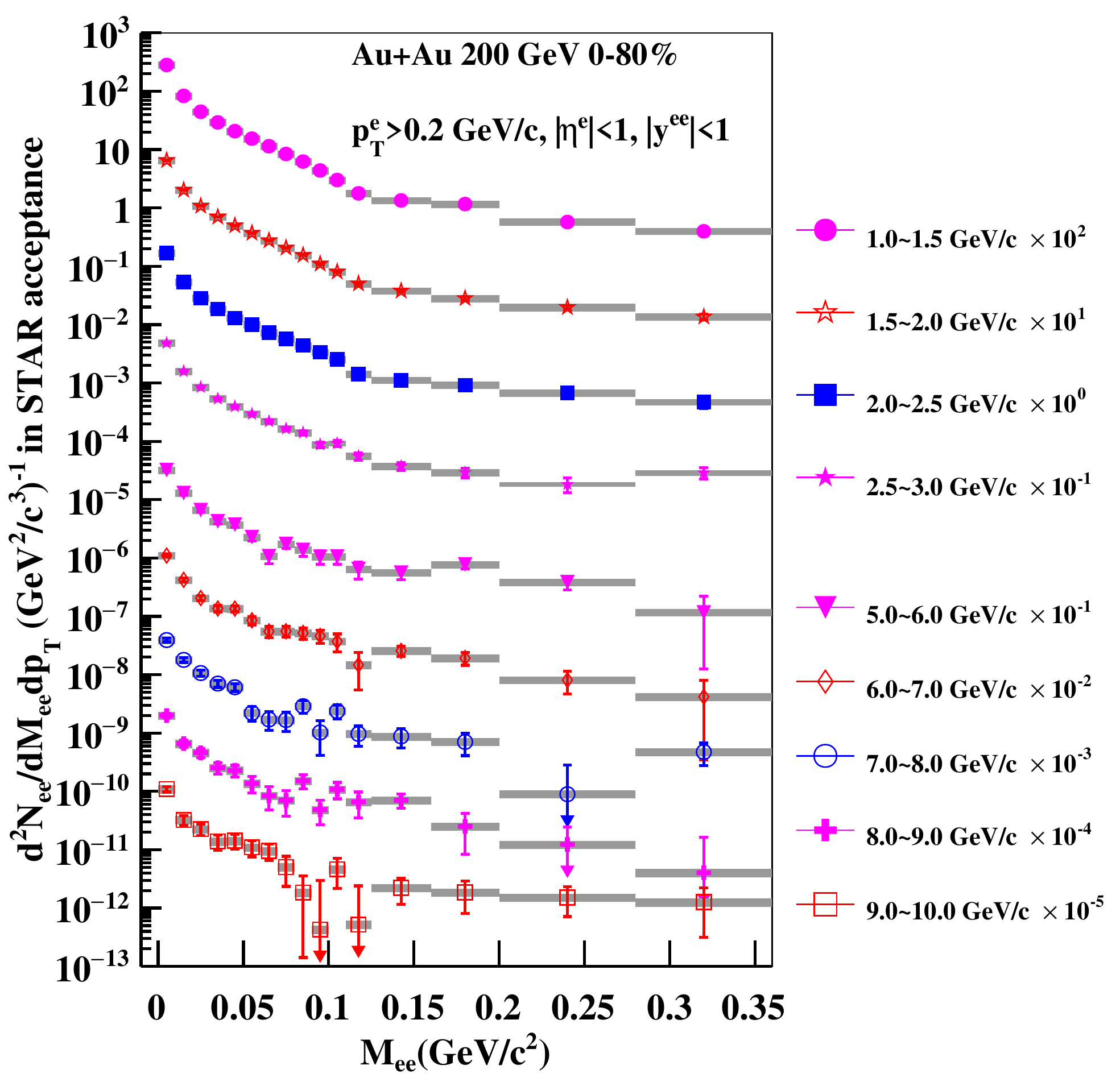}
\caption{(Color online) Dielectron invariant mass spectra in the low mass range for 0-80\% Au+Au collisions at $\sqrt{s_{_{NN}}} = 200$ GeV. The spectra in various \pt~ranges as indicated in the figure are scaled by different factors for clarity. The error bars and the shaded bands represent the statistical and systematic uncertainties, respectively.}\label{fig:continuum}
\end{center}
\end{figure}

The dielectron invariant mass spectra are obtained separately for run 10 and run 11 minimum-bias and central triggered data sets after background subtraction and efficiency correction. The analysis details for dielectron measurements from minimum-bias and central triggered events are presented in Ref.~\cite{stardielectronauauPRC}. The final results are then combined bin-by-bin according to their relative statistical uncertainties. In the mass region we are interested in, the point-to-point systematic uncertainties are dominated by the acceptance correction for the like-sign background subtraction. Due to the sector structure of the TPC, and the different bending directions of positive and negative charged particle tracks in the transverse plane, like-sign and unlike-sign pairs have different acceptances. A mixed-event technique is used to obtain the acceptance correction factor which is applied either as a function of pair invariant mass ($M_{ee}$) and $p_T$ or as a function of $M_{ee}$ only. The differences between the results from the two correction methods are taken as systematic uncertainties which are the same and correlated in run 10 and run 11. The acceptance correction factor, which is a few percent below unity at  $M_{ee}$ = 0,  increases as a function of $M_{ee}$, peaks at $M_{ee}\!=\! 0.25$ GeV/$c^2$, and reaches unity at $M_{ee} \!= \! 0.4$ GeV/$c^2$. The $M_{ee}$ and $p_T$ dependences are detailed in Ref.~\cite{stardielectronauauPRC}.
In addition, a global systematic uncertainty from the efficiency correction (14\%) is also taken into account. It is found that the systematic uncertainties of the dielectron continua in run 10 and run 11 are comparable. Therefore, the final systematic uncertainties from the combined data sets are taken as the average of those from both data sets.

For the BEMC-triggered events the dielectron pairs are formed from one electron (positron) candidate identified by the TPC and BEMC and the other positron (electron) candidate identified by the TOF and TPC. The same procedures used for the minimum-bias data set are applied to obtain the dielectron continuum signals. The details of the efficiency correction procedures are the same as reported in Ref.~\cite{stardielectronauauPRC}. Two methods are used to obtain the efficiency. In the first method we use known hadronic components as input into a Monte-Carlo simulation. In the second approach we use virtual photons as input. The resulting differences between these two methods are 4\% and are assigned as systematic uncertainties~\cite{stardielectronauauPRC}. The efficiency uncertainties in the TPC tracking, the TOF matching, and the BEMC triggering contribute to a global systematic uncertainty of 13\% for the dielectron continuum. The trigger enhancement factor is also corrected for and its uncertainty is 1\%.

 To estimate the hadron contamination effect on the dielectron continuum, we first select pure hadron samples with stringent cuts on the mass squared distributions measured from the TPC and TOF, and then  create a hadron contamination candidate pool by randomly putting in hadrons from these pure samples according to the estimated hadron contamination levels in both the total amounts and the $p_T$ differential yields. We then obtain the distributions from electron-hadron and hadron-hadron contributions utilizing the same procedures as implemented in the dielectron continuum analysis. We do not correct for the electron-hadron and hadron-hadron contributions but quote these contamination contributions as systematic uncertainties. The hadron contamination effect results in a $p_T$ dependent systematic uncertainty of (2-8)\%.  In the mass region $M_{ee}\!<0.14$ GeV/$c^2$ the uncertainty on the photon conversion rejection contributes 3\% additional systematic uncertainty for the dielectron continuum. The photon conversion rejection also removes less than 5\% of the dielectron continuum signal for $0.10\!<\!M_{ee}\!<\!0.14$ GeV/$c^2$ and this effect is corrected for~\cite{stardielectronauauPRC}. 
 
We use two approaches to estimate the efficiency correction factor from the photon conversion rejection for the dielectron continuum. In one approach, we use the $\pi^{0}$ and $\eta$ Dalitz decays as an input and get the efficiency for the dielectron signals from the Dalitz decays with the photon conversion rejection cut.  In the other approach, we use the virtual photon as an input and obtain the efficiency for the dielectrons from virtual photon decays. The resulting difference for the dielectron continuum in the efficiency correction difference (3\%) from the two approaches is quoted as part of the systematic uncertainties.
The total systematic uncertainty of measured yields is 15-16\% which is independent of $M_{ee}$ and has a slight $p_T$ dependence.

The dielectron invariant mass spectrum in this analysis is constructed within the STAR acceptance ($p_{T}^{e}>0.2$ GeV/$c$,$~|\eta^{e}|<1,~|y^{ee}|<1$) and corrected for efficiency, where $p_{T}^{e}$ is the electron \pt, $\eta^{e}$ is the electron pseudo-rapidity, and $y^{ee}$ is the rapidity of electron-positron pairs.  The dielectron invariant mass spectra in different dielectron \pt~ranges are shown in Fig.~\ref{fig:continuum}. The results for $p_{T}<3$ GeV/$c$ are the combined results from run 10 and run 11 minimum-bias and central triggered data as reported in~\cite{stardielectronauauPRC}. The results for $p_{T}>5$ GeV/$c$ are from the BEMC-triggered data. The limitation of the dielectron \pt~reach in these two data sets is due to a large hadron contamination for electrons at $p_T^{e}\!>\!2$ GeV/$c$ in minimum-bias and central triggered data and a low trigger efficiency for electrons at $p_T^{e}\!<\!4.5$ GeV/$c$ in the BEMC-triggered data.

The relation between real photon yield and the associated \epem~pair production can be described as in \mbox{Eq.~\ref{eq:dvp2ee}}~\cite{DVP2ee,Landsberg},
\begin{equation}\label{eq:dvp2ee}
\frac{d^{2}N_{ee}}{dM_{ee}dp_T} = \frac{2\alpha}{3\pi}\frac{1}{M_{ee}}L(M_{ee})S(M_{ee},p_T)\frac{dN_{\gamma}}{dp_T}.
\end{equation}
Here, $L(M_{ee})= \sqrt{1-\frac{4m_{e}^{2}}{M_{ee}^{2}}}(1+\frac{2m_{e}^{2}}{M_{ee}^{2}})$, $\alpha$ is the fine structure constant, $M_{ee}$ is the \epem~pair mass, $m_e$ is the electron mass, and $S(M_{ee}, p_T)$ is a process-dependent factor accounting for differences between real and virtual photon production. We adopted the same assumption as in Ref.~\cite{thermalphoton}, namely that the factor $S(M_{ee}, p_T)$ is approximately 1 for $M_{ee}<0.3$ GeV/$c^{2}$, $p_{T}>1$ GeV/$c$. The uncertainty associated with this assumption is expected~\cite{galerappprivate} to be insignificant compared to the uncertainty in the data. Therefore, we do not assign any systematic uncertainty for this assumption. For $M_{ee}>>m_e$, the factor $L(M_{ee})$ is also unity. Thus the relation becomes
\begin{equation}\label{eq:3}
\frac{d^{2}N_{ee}}{dM_{ee}dp_T} \approx \frac{2\alpha}{3\pi}\frac{1}{M_{ee}}\frac{dN_{\gamma}}{dp_T}.
\end{equation}
If there is direct real photon production in a given $p_T$ bin, then there should be a corresponding electron pair production which behaves like $1/M_{ee}$ in the same $p_T$ bin, as indicated by Eq.~\ref{eq:3}. Thus, the direct real photon production can be derived from the yield of the excess dielectron pairs.

\begin{figure}
\begin{center}
\includegraphics*[width=0.47\textwidth]{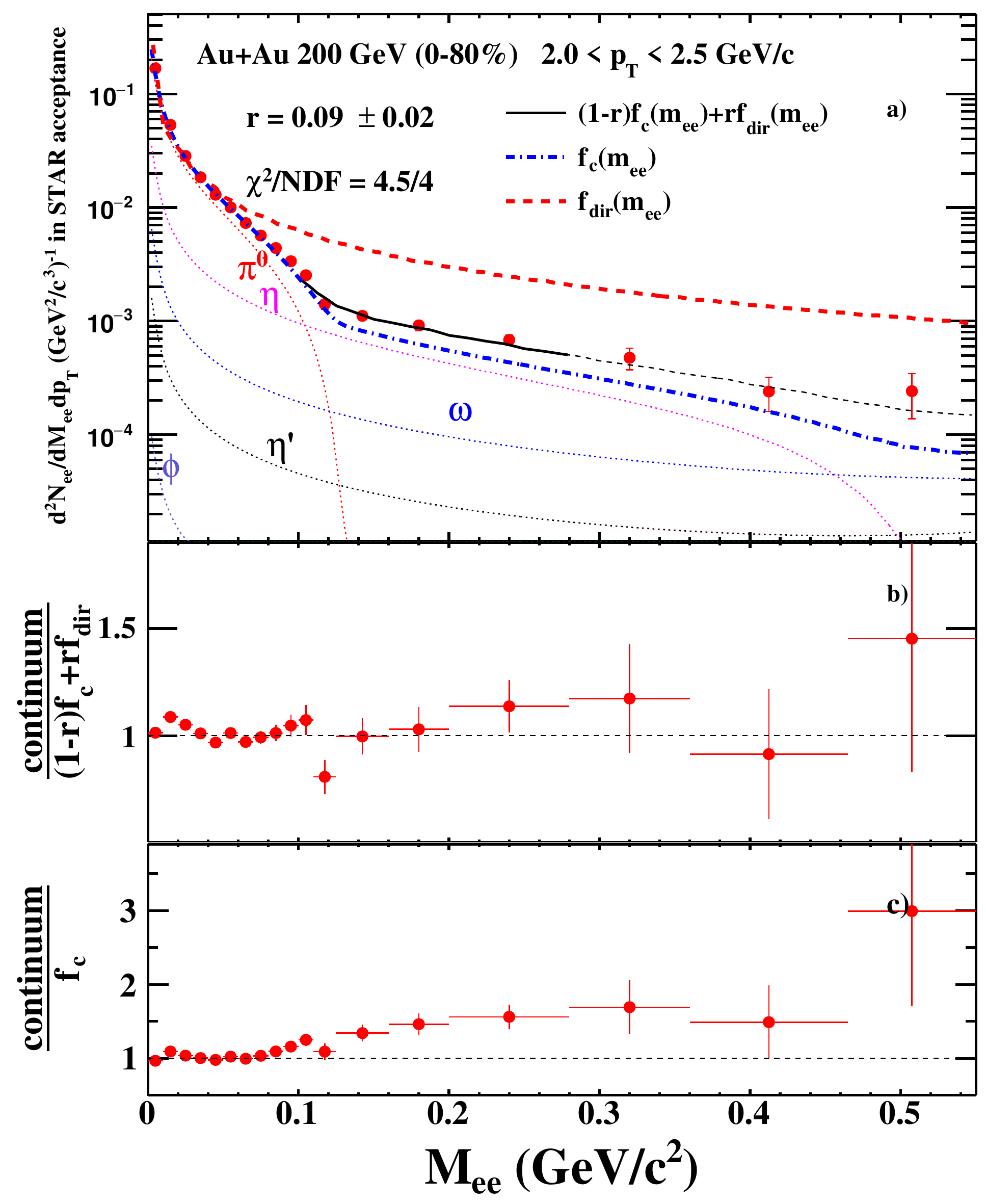}
\caption{(Color online) Panel (a): The two-component fitting function results for the Au+Au dielectron spectra at $2.0<p_{T}<2.5$ GeV/$c$. The uncertainties in the dielectron mass spectrum are the quadrature sum of statistical and point-to-point systematic uncertainties. The dot-dashed and dashed lines represent the normalized cocktail and internal conversion from direct photons, respectively. The solid line is the fit to the data in the range $0.10<M_{ee}<0.28$ GeV/$c^{2}$. The light dashed-line is the extrapolation of the fit function outside the fit range. The dotted lines represent different cocktail components. The $c\bar{c}$ contribution is omitted for clarity. Panel (b): The data divided by the fit model as a function of $M_{ee}$. Panel (c): The data divided by the cocktail component as a function of $M_{ee}$.}\label{fig:fitting}
\end{center}
\end{figure}

The direct photon yields are extracted by fitting the dielectron invariant mass spectra in the low mass region with two components. In the two-component fitting function $(1-r)f_{\text{cocktail}}+rf_{\text{dir}}$, $f_{\text{cocktail}}$ is the shape of the normalized hadronic cocktail mass distribution within the STAR acceptance, $f_{\text{dir}}$ is the shape of the normalized, internal conversion mass distribution from direct photons within the STAR acceptance, and $r$ is a fitting parameter. The first term $(1-r)f_{\text{cocktail}}$ in the fitting function represents the background, namely the contribution from known hadronic sources. These include $\pi^{0}$, $\eta$, and $\eta^{\prime}$
Dalitz decays: $\pi^{0}\rightarrow \gamma e^{+}e^{-}$, $\eta
\rightarrow \gamma e^{+}e^{-}$, and $\eta^{\prime}\rightarrow
\gamma e^{+}e^{-}$; vector meson decays: $\omega \rightarrow \pi^{0}e^{+}e^{-}$ and $\phi
\rightarrow \eta e^{+}e^{-}$; and
heavy-flavor hadron semi-leptonic decays: $c\bar{c} \rightarrow
e^{+}e^{-}$. Among those, $\pi^{0}$ and $\eta$ Dalitz decays are dominant contributions. The second term $rf_{\text{dir}}$ represents the signal, i.e. direct photon internal conversion. The cocktail components are the same as in Ref.~\cite{stardielectronauauPRC}. We normalize both $f_{\text{cocktail}}$ and $f_{\text{dir}}$ to data points for $M_{ee}\!<0.03$ GeV/$c^2$, separately. In this mass region the shapes of $f_{\text{cocktail}}$ and $f_{\text{dir}}$ are identical, thus the fitting function in this mass region is independent of $r$. The parameter $r$ can be interpreted as the ratio of direct photon to inclusive photon yields. The range for the two-component fit to data is $0.10\!<M_{ee}\!<0.28$ GeV/$c^2$.

Figure~\ref{fig:fitting} shows an example of the two-component fit for $2.0<p_{T}<2.5$ GeV/$c$. We note that there is a small peak structure at $M_{ee}=0.02$ GeV/$c^2$ in the ratio plots as indicated in panels (b) and (c). This peak could be due to an imperfect description of the material budget in the photon conversion simulations. To estimate this effect on our results we varied the range for $f_{\text{cocktail}}$ and $f_{\text{dir}}$ to be normalized to the data from $M_{ee}\!<0.03$ GeV/$c^2$ to $M_{ee}\!<0.05$ GeV/$c^2$. The resulting difference for the virtual photon yields compared to the default case is $(0.2-1.0)$\% and is included as part of the systematic uncertainties.

With the $r$ value derived for each $p_T$ bin, one can obtain the direct virtual photon invariant yield $\frac{d^{2}N_{\gamma}^{dir}(p_T)}{2\pi p_{T}dp_{T}dy}$ as a function of $p_T$. The detailed methodology can be found in Ref.~\cite{DVP_PHENIX_detailed}. From \mbox{Eq.~\ref{eq:3}}, the direct virtual photon term in two-component fit can be written as \mbox{Eq.~\ref{eq:4}}. Then the direct virtual photon invariant yield as a function of $p_T$ can be written in \mbox{Eq.~\ref{eq:dvpyield}}:
\begin{equation}\label{eq:4}
\frac{2\alpha dN_{\gamma}^{dir}(p_{T})}{3\pi M_{ee} dp_{T}}=r  F_{dir} \frac{1}{M_{ee}},
\end{equation}

\begin{equation}\label{eq:dvpyield}
\frac{d^{2}N_{\gamma}^{dir}(p_T)}{2\pi p_{T}dp_{T}dy}=\frac{3rF_{dir}} {4\alpha p_{T}dy}=r \frac{d^{2}N_{\gamma}^{inc}(p_T)}{2\pi p_{T}dp_{T}dy},
\end{equation} in which $\frac{d^{2}N_{\gamma}^{dir}(p_T)}{2\pi p_{T}dp_{T}dy}$, $\frac{d^{2}N_{\gamma}^{inc}(p_T)}{2\pi p_{T}dp_{T}dy}$, $dp_T$, $dy$, and $F_{dir}$ are the direct photon invariant yield, inclusive photon invariant yield, $p_T$ bin width, rapidity bin width, and $f_{dir}$ normalization factor, respectively.

\begin{table*}\caption{Sources and their contributions to the relative systematic uncertainties for direct virtual photon yields in different centralities. The $p_T$ dependent uncertainties for each source are listed as a range. The 15\% overall systematic uncertainty, labelled as ``global", is dominated by the efficiency correction and is $p_T$ independent. Contributions from $\eta'$ and $\omega$ are negligible. The difference between the dielectron continuum distributions in run 10 and run 11 results in an overall systematic uncertainty for each centrality and is labelled as ``RunDiff." The total systematic uncertainties are the quadratic sums of the different contributions.\label{tabI}}
{\centering
\begin{tabular}{c|c|c|c|c|c} \hline\hline
 source & centrality 0-80\% & centrality 0-20\% & centrality 20-40\% & centrality 40-60\% & centrality 60-80\% \\ \hline
\hspace{0.16in}fit range&  14\% &13\% & 15\% &9\% &16\%\\
$\pi^{0}/\eta$ & 2-43\% & 2-31\% &1-35\% &2-71\% &1-70\%\\
$c\bar{c}$ & 0-6\% &0-4\% &0-4\% &0-6\% &0-5\%\\
global & 15\% & 15\%& 15\%&15\%&15\%\\
normalization& 0.2\% & 0.2\% & 0.1\% & 0.2\% & 0.1\%\\
RunDiff & 2.2\% & 2.7\%& 0.8\% &0.5\% &1.2\%\\ \hline
total & 20-48\% & 19-37\%& 21-41\%&17-73\%&21-74\%\\
\hline \hline
\end{tabular}
}
\end{table*}

We fit the dielectron continuum with statistical and systematic uncertainties. The fit errors contribute to the statistical uncertainties for the direct virtual photon yields.
Systematic uncertainties for direct virtual photon yields are mainly from the two-component fit, which is dominated by the uncertainties in the cocktail and the fit range. The fitting range uncertainty is estimated by taking the full difference between the results obtained by varying the fit range from 0.10-0.28 GeV/$c^2$, to 0.08-0.28, and to 0.12-0.28 GeV/$c^2$. Extending the fit range to $M_{ee}~<~$0.36 GeV/$c^2$ results in a negligible systematic uncertainty. These ranges are selected based on three criteria: $p_T/M_{ee}\!\gg1$, $M_{ee}$ far enough away from the normalization region ($M_{ee}\!<\!0.03$ GeV/$c$), and availability of the data. The normalization range for $f_{\text{cocktail}}$ and $f_{\text{dir}}$ in the two-component fit contributes less than 1\% systematic uncertainty, as explained in the previous sections and shown in Table~\ref{tabI}. For the cocktail the uncertainties in the total cross sections for $\pi$~\cite{AuAuPID} and charm ($c\bar{c}$)~\cite{charm} are 8\% and 45\%, respectively, independent of $p_T$. For the default $\eta$ $p_T$ spectrum, a Tsallis blast-wave model prediction, with the freeze-out parameters obtained by fitting other hadrons simultaneously, is used. We then obtain the ratio of $\eta$ over $\pi$ as a function of $p_T$ and match it to the $\eta/\pi$ ratio value measured by PHENIX at $p_T$=5 GeV/$c$~\cite{phenixeta,thermalphoton}. For the systematic uncertainty study, we vary the $\eta/\pi$ ratio by 13\% as used in Ref.~\cite{phenixeta,thermalphoton}. The uncertainties in the cross sections for combined $\pi$ and $\eta$ and $c\bar{c}$, mentioned above, result in uncertainties of 2-43\% and 0-6\%, respectively, decreasing as a function of $p_T$ for the direct virtual photon yields for 0-80\% Au+Au collisions. We note that the PHENIX Collaboration does not use a Tsallis blast-wave model prediction to constrain the $\eta$ $p_T$ spectrum at low $p_T$ but use a so-called transverse mass ($m_T$) scaling~\cite{thermalphoton}. In our analysis, we also obtain the direct  virtual photon yields using the $m_T$ scaling and compare them to the default results based on a Tsallis blast-wave model prediction.
Contributions from $\eta'$ and $\omega$ are negligible for the hadronic cocktail, resulting in a negligible contribution for the systematic uncertainties. In addition, the 15\% overall systematic uncertainty, dominated by the efficiency correction to the dielectron continuum, is $p_T$ independent for the direct virtual photon yields and does not affect the ratio of direct photon to inclusive photon yields. Table~\ref{tabI} lists sources and their contributions to the systematic uncertainties for the direct virtual photon yields in different centralities. The total systematic uncertainties are the quadratic sums of the different contributions.

\section{Results}
\label{results}
\begin{figure}
\begin{center}
\includegraphics*[width=0.47\textwidth]{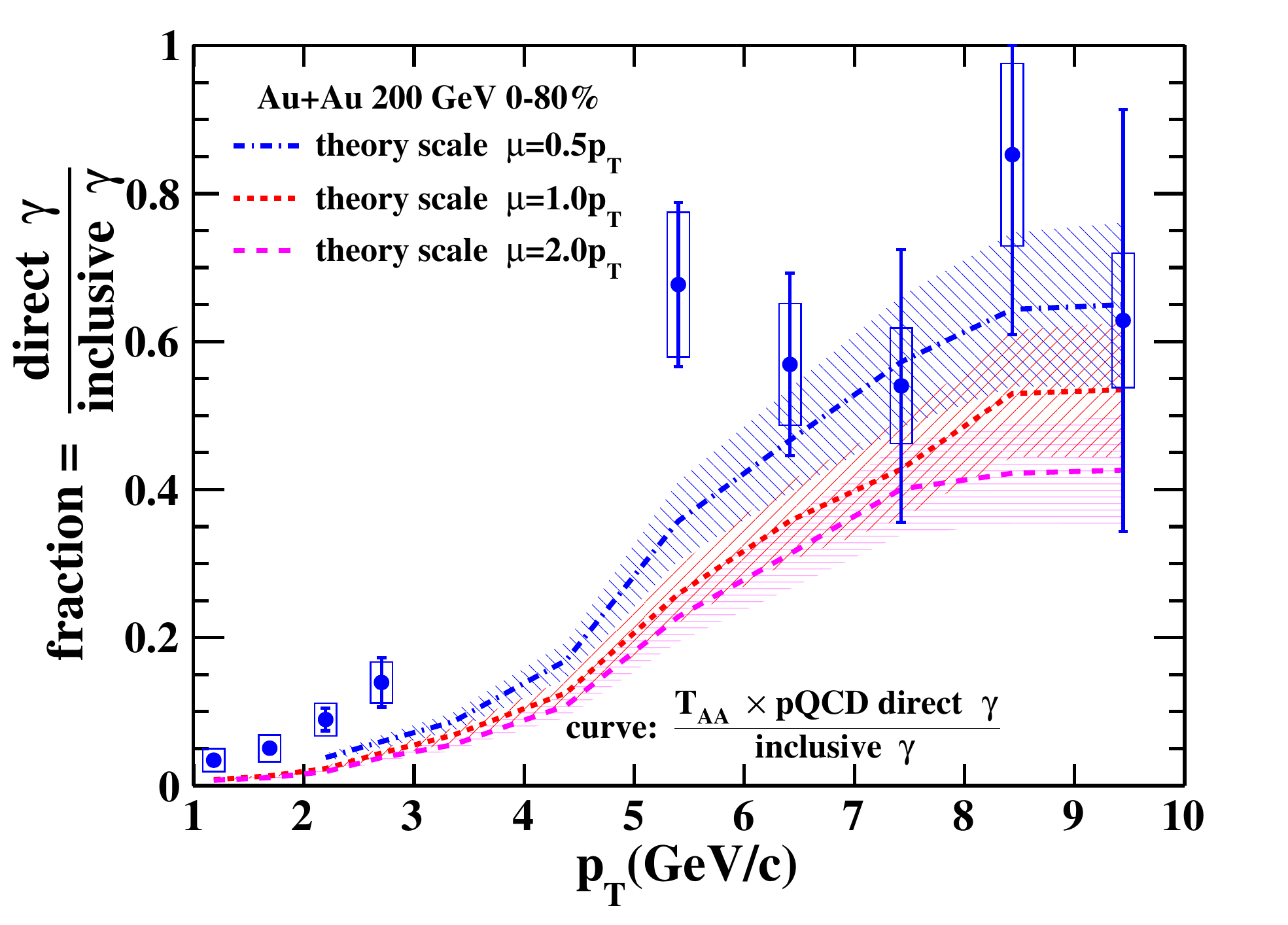}
\caption{(Color online) The ratio of direct photon to inclusive photon yields compared with the ratio of $T_{AA}$ scaled NLO pQCD predictions to inclusive photon yields for 0-80\% Au+Au collisions at $\sqrt{s_{_{NN}}} = 200$ GeV. The data points for $1\!<\!p_T\!<\!3$ GeV/$c$ and $5\!<\!p_T\!<\!10$ GeV/$c$ are from minimum-bias data and calorimeter-triggered data, respectively. The three curves correspond to pQCD calculations with different renormalization ($\mu_{R}$) and factorization scales ($\mu_{F}$), assuming $\mu_{R} = \mu_{F} = \mu$. The error bars and the boxes represent the statistical and systematic uncertainties, respectively. The shaded bands on the curves represent the systematic uncertainties for inclusive photon measurements, which are about 15\%.}\label{fig:fraction}
\end{center}
\end{figure}

Figure~\ref{fig:fraction} shows the $r$ value, the ratio of direct photon to inclusive photon yields compared with the ratio of $T_{AA}$ scaled Next-to-Leading-Order (NLO) perturbative QCD (pQCD) predictions to inclusive photon yields as a function of $p_{T}$. The curves represent $T_{AA}\frac{d^{2}\sigma_{\gamma}^{\text{NLO}}(p_{T})}{2\pi p_T dp_T dy}/\frac{d^2N_{\gamma}^{\text{inc}}(p_{T})}{2\pi p_T dp_T dy}$ showing the scale dependence of the theory~\cite{NLOpQCDcurve} in which $T_{AA}$ is the nuclear overlap factor, $\frac{d^{2}\sigma_{\gamma}^{\text{NLO}}(p_{T})}{2\pi p_T dp_T dy}$ is the \pt-differential invariant cross section for direct photons obtained from Ref.~\cite{DVP_PHENIX_pp}, and $\frac{d^2N_{\gamma}^{\text{inc}}(p_{T})}{2\pi p_T dp_T dy}$ is the inclusive photon \pt-differential invariant yield. The data show consistency with NLO pQCD calculations within uncertainties at $p_{T}>6$ GeV/$c$. A clear enhancement in data compared to the calculation for $1<p_{T}<3$ GeV/$c$ is observed. The data point at $p_{T}\!=\!5.5$ GeV/$c$ is about 1.8$\sigma$ higher than the calculation.

\begin{figure}
\begin{center}
\includegraphics*[width=0.47\textwidth]{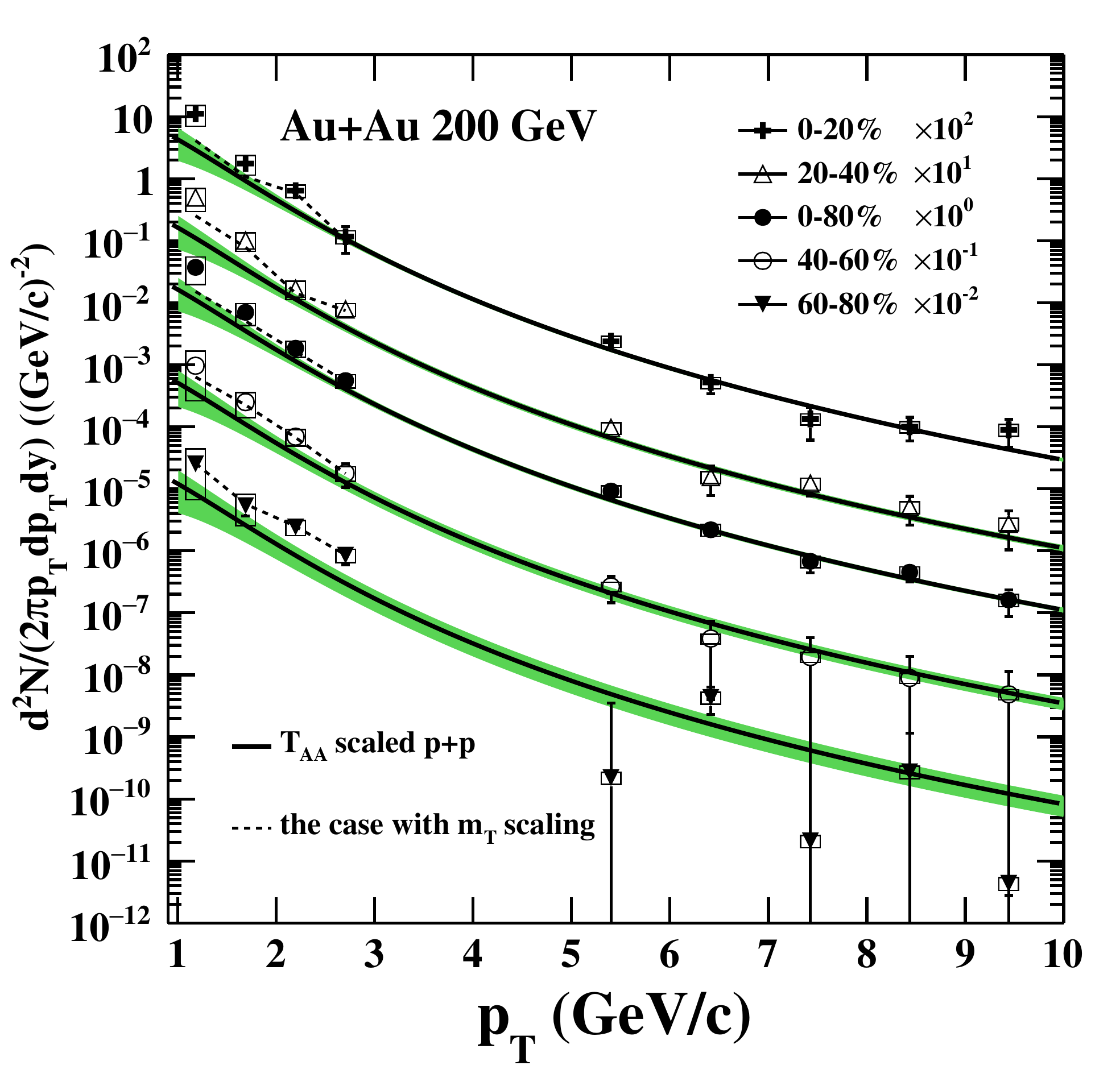}
\caption{(Color online) Centrality dependence of the direct photon invariant yields as a function of $p_T$ in Au+Au collisions at $\sqrt{s_{_{NN}}} = 200$ GeV. The solid curves represent a power-law fit to PHENIX 200 GeV $p+p$ results~\cite{Mizuno:14,DVP_PHENIX_pp}, scaled by $T_{AA}$. The bands on the curves represent the uncertainties in the parameterization and in $T_{AA}$. The dashed lines represent the direct photon invariant yields when we use the $m_T$ scaling to constrain the $\eta/\pi$ ratio.  See the text for detailed discussions. The error bars and boxes represent the statistical and systematic uncertainties, respectively.}\label{fig:yield}
\end{center}
\end{figure}

Figure~\ref{fig:yield} shows centrality dependence of the invariant yields of direct photons in Au+Au collisions at $\sqrt{s_{_{NN}}} = 200$ GeV.  The $p+p$ results are parameterized by a power-law function~\cite{DVP_PHENIX_pp}, the same one as used in Ref.~\cite{Mizuno:14}. The parameterized distribution is then scaled by $T_{AA}$, and compared to the Au+Au results in different centralities, as shown by the solid curves. The $T_{AA}$ values calculated from a Glauber model for 0-20\%, 20-40\%, 0-80\%, 40-60\%, and 60-80\% Au+Au collisions at $\sqrt{s_{_{NN}}} = 200$ GeV are $(766 \pm 28) / 42$ $\mathrm{mb}$, $(291 \pm 30)/ 42$ $\mathrm{mb}$, $(292 \pm 20) / 42$ $\mathrm{mb}$, $(91 \pm 20)/ 42 $ $\mathrm{mb}$, and $(22 \pm 8)/ 42$ $\mathrm{mb}$, respectively. For $1<p_{T}<3$ GeV/$c$, the Au+Au results are higher than $T_{AA}$ scaled $p+p$ results, while at $p_{T}>6$ GeV/$c$ the Au+Au yield is consistent with the scaled $p+p$ expectation. We note that for $1<p_{T}<2$ GeV/$c$, the data points in 40-60\%  and 60-80\% Au+Au collisions have larger uncertainties and are also consistent with the scaled $p+p$ expectations. Also shown in Fig.~\ref{fig:yield} are the direct virtual photon yields in different centralities when we use the $m_T$ scaling to constrain the $\eta/\pi$ ratio. We note that the result based on the $m_T$ scaling differs more from the default case in central collisions while in 60-80\% peripheral collisions the result based on the $m_T$ scaling is identical to the default case since the flow effect is negligible on the $\eta$ $p_T$ spectrum in peripheral collisions.

\begin{figure}
\begin{center}
\includegraphics*[width=0.47\textwidth]{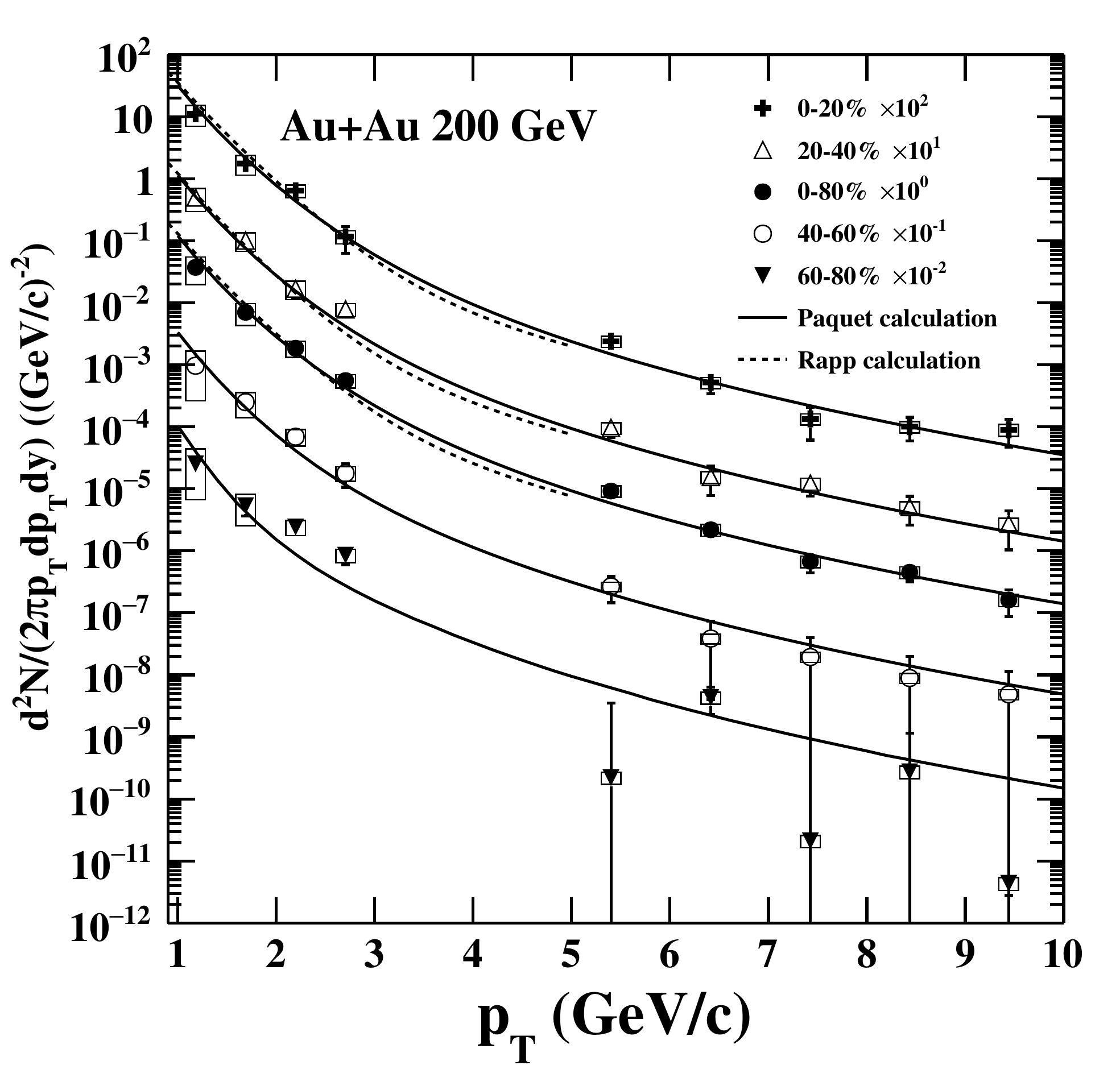}
\caption{The direct photon invariant yields as a function of $p_T$ in Au+Au collisions at $\sqrt{s_{_{NN}}} = 200$ GeV compared to model predictions from  Rapp et al.~\cite{rapp:11,private1} and Paquet et al.~\cite{McGill:15} . The statistical and systematic uncertainties are shown by the bars and boxes, respectively. }\label{fig:yieldcompare}
\end{center}
\end{figure}

A comparison between STAR Au+Au data and model calculations from  Rapp $et~al.$~\cite{rapp:11,private1} and Paquet $et~al.$~\cite{McGill:15} is shown in Fig.~\ref{fig:yieldcompare}.  For the direct photon production both models include the contributions from QGP thermal radiation, in-medium $\rho$ meson and other mesonic interactions in the hadronic gas, and primordial contributions from the initial hard parton scattering. In Refs.~\cite{rapp:11,private1} an elliptic thermal fireball evolution is employed for the bulk medium. Non-thermal primordial photons from $N_{bin}$ collisions are estimated from either a pQCD-motivated $x_T$-scaling ansatz or a parameterization
of PHENIX $p+p$ data. The sum of the thermal medium and primordial contributions for the former case is shown in Fig.~\ref{fig:yieldcompare}. Using a parameterization of PHENIX $p+p$ reference data would lead to slightly higher direct photon yields. In addition, a (2+1)-D hydrodynamic evolution (beam-direction independent) is employed for the bulk medium by Rapp $et~al.$ and the results are consistent with those from the fireball evolution. In Ref.~\cite{McGill:15} a (2+1)-D hydrodynamic evolution is employed for the bulk medium.
Comparison of the model and data shows that in the \pt~range 1-3 GeV/$c$ the dominant sources are from thermal radiation while, as \pt~increases to 5-6 GeV/$c$, the initial hard-parton scattering becomes dominant. The comparison shows consistency between both model calculations and our measurement within uncertainties for all the other centralities except 60-80\% centrality, where hydrodynamic calculations might not be applicable. We note that in the centrality determination there is a large uncertainty in peripheral collisions, as seen in the $N_{bin}$ uncertainty.

\begin{figure}
\begin{center}
\includegraphics*[width=0.47\textwidth]{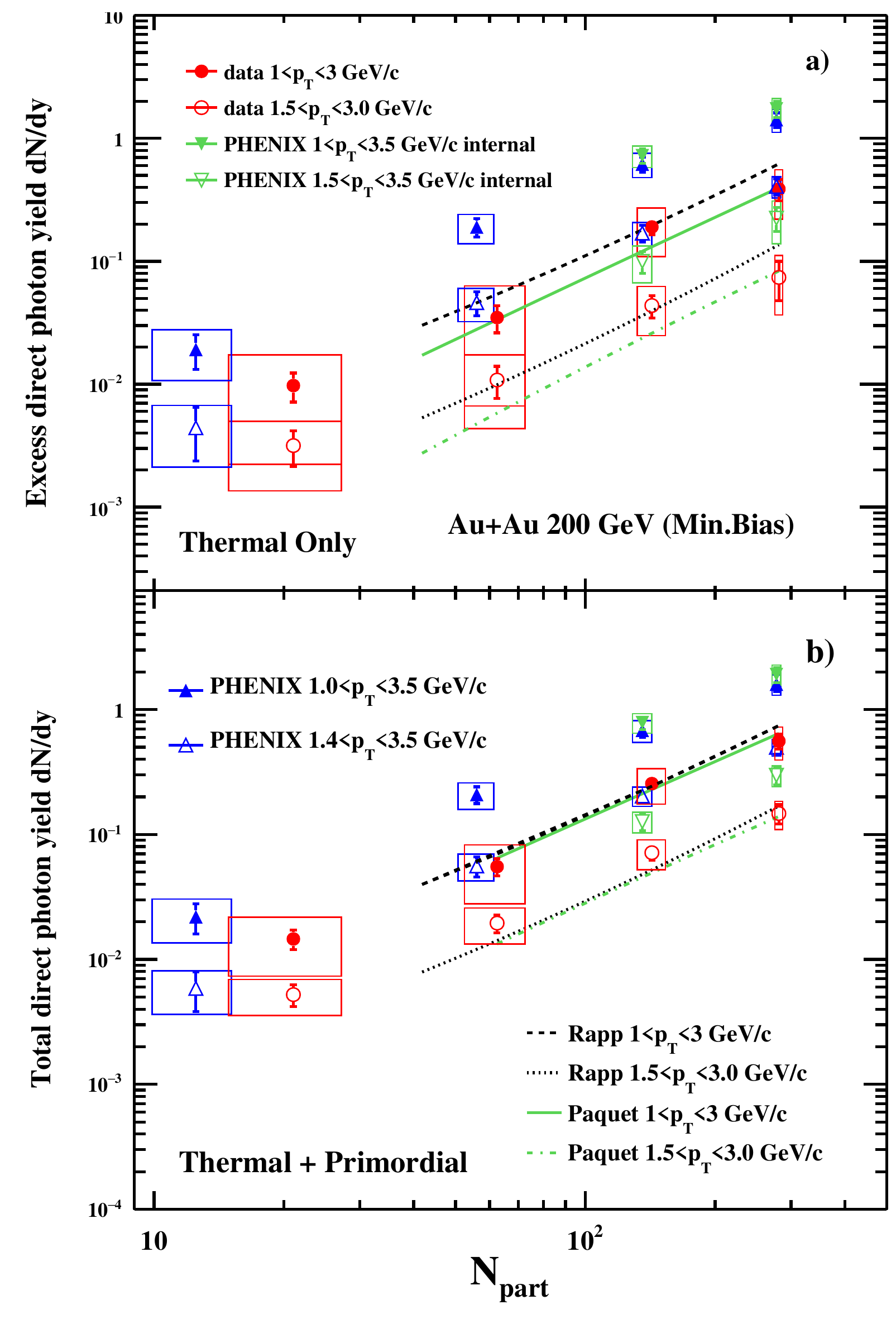}
\caption{(Color online) The excess [panel (a)] and total [panel (b)] direct photon yields in different $p_T$ ranges as a function of the number of participating nucleons ($N_{part}$) from STAR (circles) and PHENIX (triangles) in Au+Au collisions at $\sqrt{s_{_{NN}}} = 200$ GeV.  The down-pointing triangles represent the results from the internal conversion method~\cite{thermalphoton} while the up-pointing triangles represent the results from Ref.~\cite{Mizuno:14}. Model predictions from  Rapp et al.~\cite{rapp:11,private1} and Paquet et al.~\cite{McGill:15}  are also shown for the excess  [panel (a)] and total [panel (b)] direct photon yields. The statistical and systematic uncertainties are shown by the bars and boxes, respectively. } \label{fig6}
\end{center}
\end{figure}

We integrate the direct virtual photon yields in different $p_T$ ranges,  study their centrality dependences, and compare the data from STAR and PHENIX as well as the theoretical model calculations described above. For the STAR measurements, we use two $p_T$ bins: 1-3 GeV/$c$ and 1.5-3 GeV/$c$. For the PHENIX measurements in Ref.~\cite{thermalphoton}, the same $p_T$ bins are used for 0-20\% and 20-40\% centrality bins.
For the PHENIX measurements in Ref.~\cite{Mizuno:14}, we use 1-3.5 GeV/$c$ and 1.4-3.5 GeV/$c$. Different ranges are selected due to the availability of the data. Theoretical model calculations show that the contribution of the yield in the $p_T$ range 3-3.5 GeV/c is 0.4\% to the yield in the range of  1-3.5 GeV/$c$ . The contributions in the $p_T$ ranges 3-3.5 GeV/$c$ and 1.4-1.5 GeV/$c$ are 25\% to the yield in the range 1.4-3.5 GeV/$c$.
Figure~\ref{fig6} shows the comparison of the data and the theoretical model calculations~\cite{rapp:11,private1,McGill:15}. Panel (a) presents the excess yield, which is the direct photon yield with the $T_{AA}$ scaled $p+p$ contribution subtracted, in comparison with the thermal component contributions in the model calculations.  Since the $p+p$ references have a large uncertainty, we also compare the total direct photon yield to the sum of thermal and primordial contributions in the models, as shown in panel (b).
The comparisons indicate that our measurements of the excess and total yields are systematically lower than the PHENIX results in 0-20\%, 20-40\%, and 40-60\% centrality bins. The model calculations are consistent with our measurements within uncertainties. We note that the two model calculations give similar total yields but different thermal contributions. For the comparisons between data and model calculations, the $\chi^2/NDF$ and p-value are listed in Table~\ref{tabII}. Note that the models with the same physics ingredients~\cite{ralf,ralf:08,PHSD:12,USTC:12} describe the dilepton measurements~\cite{stardielectronauau,stardielectronauauPRC,phenixdielectronauau,stardielectronauauPLB,na60dimuon,CERES}. Models with additional, new physics ingredients~\cite{photonNewTheo}, which attempt to describe the PHENIX photon data, should be compared to the world-wide photon and dilepton data for a consistency check. In the future, more precise measurements of direct photons in both heavy ion and p+p collisions are needed to further distinguish between different model calculations.

\begin{table}\caption{The $\chi^2/NDF$ and p-value in central and mid-central collisions between data and model calculations for $1\!<p_T\!<\!3$ GeV/$c$.\label{tabII}}
{\centering
\begin{tabular}{c|c|c} \hline\hline
 Comparison & $\chi^2/NDF$ & p-value  \\ \hline
\hspace{0.16in}Excess yield&  & \\
STAR data to Rapp & 2.1/2 & 0.35 \\
STAR data to Paquet   &0.49/2 & 0.78\\
PHENIX internal conversion~\cite{thermalphoton} to Rapp & 15/1 &1.1e-04 \\
PHENIX internal conversion~\cite{thermalphoton} to Paquet & 20/1 & 7.7e-06\\
PHENIX data~\cite{Mizuno:14} to Rapp & 17/2 & 2.0e-04 \\
PHENIX data~\cite{Mizuno:14} to Paquet & 24/2 & 6.1e-06\\ \hline
Total yield&  & \\
STAR data to Rapp & 1.4/2 & 0.50 \\
STAR data to Paquet   &0.55/2 & 0.76\\
PHENIX internal conversion~\cite{thermalphoton} to Rapp & 16/1 &6.3e-05 \\
PHENIX internal conversion~\cite{thermalphoton} to Paquet & 18/1 & 2.2e-05\\
PHENIX data~\cite{Mizuno:14} to Rapp & 19/2 &7.5e-05 \\
PHENIX data~\cite{Mizuno:14} to Paquet & 21/2 & 2.7e-05\\ \hline
\hline
\end{tabular}
}
\end{table}


\section{Conclusions}
\label{conclusion}
We measured \epem~spectra and inferred direct photon production in Au+Au collisions at STAR at $\sqrt{s_{NN}}=200$ GeV. The direct photon measurement based on the virtual photon method is extended to \pt~of 5-10 GeV/$c$. In the \pt~range 1-3 GeV/$c$ the direct photon invariant yield shows a clear excess in 0-20\% and 20-40\% central Au+Au over the $T_{AA}$ scaled $p+p$ results. In the \pt~range above 6 GeV/$c$ there is no clear enhancement observed for all the centralities. Model predictions which include the contributions from thermal radiation and initial hard-processes are consistent with our direct photon yield within uncertainties in 0-20\%, 20-40\%, and 40-60\% collisions. In 60-80\% centrality bin, the model calculation results are  systematically lower than our data for $2\!<\!p_T\!<\! 3$ GeV/$c$.





\vspace{0.2in}
{\bf Acknowledgments}
\vspace{0.1in}

We thank the RHIC Operations Group and RCF at BNL, the NERSC Center at LBNL, the KISTI Center in
Korea, and the Open Science Grid consortium for providing resources and support. This work was
supported in part by the Office of Nuclear Physics within the U.S. DOE Office of Science,
the U.S. NSF, the Ministry of Education and Science of the Russian Federation, NSFC, CAS,
MoST and MoE of China, the National Research Foundation of Korea, NCKU (Taiwan),
GA and MSMT of the Czech Republic, FIAS of Germany, DAE, DST, and UGC of India, the National
Science Centre of Poland, National Research Foundation, the Ministry of Science, Education and
Sports of the Republic of Croatia, and RosAtom of Russia. We thank C. Gale, J. Paquet, R. Rapp, C. Shen, and H. van Hees for valuable discussions and for providing the theoretical calculations.

\end{document}